\documentclass[12pt]{iopart}
\usepackage{epsfig}
\usepackage{bm}

\newcommand{\be}{\begin{eqnarray}}
\newcommand{\ee}{\end{eqnarray}}
\newcommand{\bmat}{\left(\begin{array}}
\newcommand{\emat}{\end{array}\right)}
\newcommand{\no}{\nonumber}
\newcommand{\diff}{\mathrm d}

\begin{document}

\title[Shortcuts to adiabaticity applied to nonequilibrium entropy production]
{Shortcuts to adiabaticity applied to nonequilibrium entropy production: An information geometry viewpoint}
\author{Kazutaka Takahashi}
\address{Department of Physics, Tokyo Institute of Technology, Tokyo 152-8551, Japan}

\vspace{10pt}
\begin{indented}
\item[]\today
\end{indented}

\begin{abstract}
We apply the method of shortcuts to adiabaticity to nonequilibrium systems.
For unitary dynamics, the system Hamiltonian is separated into two parts.
One of them defines the adiabatic states for the state to follow and 
the nonadiabatic transitions are prevented by the other part.
This property is implemented to the nonequilibrium entropy production 
and we find that the entropy is separated into two parts.
The separation represents the Pythagorean theorem 
for the Kullback--Leibler divergence and 
an information-geometric interpretation is obtained.
We also study a lower bound of the entropy, which 
is applied to derive a trade-off relation between time, entropy and 
state distance.
\end{abstract}

%%%%%%%%%%%%%%%%%%%%%%%%%%%%%%%%%%%%%%%%%%%%%%%%%%%%%%%%%%%%%%%%%%%%%%%%%%%
\section{Introduction}

Understanding nonequilibrium properties of dynamical systems 
is a fascinating topic in physics and has been studied intensively.
The fluctuations of thermodynamic functions 
are considered to be key properties, and 
the Jarzynski equality~\cite{Jarzynski1, Jarzynski2}
and the fluctuation theorem~\cite{ECM, GC} play the prominent roles.
The nonequilibrium entropy production is one of quantities to measure
nonequilibrium properties of the system and has been studied 
in many contexts.
Especially, knowing the lower bound is an important task since it determines 
irreversibility, dissipation properties, efficiency 
and so on~\cite{AN1, AN2, KPB, VJ, DL1, DL2, DPZ}.

Thermally-isolated quantum systems can be 
treated by the unitary dynamics of the Schr\"odinger equation.
In this paper we characterize the dynamics by 
the method of shortcuts to adiabaticity (STA).
This method enables us to achieve an adiabatic dynamics in a finite time.
To prevent the nonadiabatic transitions, we introduce an additional term 
called the counterdiabatic term.
The fundamental idea was pointed out using a simple two-level Hamiltonian 
in~\cite{EZAN} 
and the general formulation was developed 
in several works~\cite{DR1, DR2, Berry, CRSCGM}.
Since then, the method has been intensively studied in various way~\cite{STA}.
We can find applications to simple systems~\cite{CLRGM, MCILR}, 
scale-invariant systems~\cite{delCampo}, 
many-body systems~\cite{dCRZ, Takahashi1, OT1}, and 
classical systems~\cite{Jarzynski3, DJdC, PJ, OT2} and so on.
The method can also be implemented experimentally
to several systems~\cite{SSVL, SSCVL, Betal, Zetal}.
It is also expected to be applied to quantum computations 
such as the quantum annealing.

It should be stressed that STA is applied to any dynamical systems.
STA is useful not only to control the system 
but also to describe general unitary dynamics.
As we describe below, the system Hamiltonian is 
separated into two parts, $\hat{H}(t)=\hat{H}_0(t)+\hat{H}_1(t)$, 
and the state satisfying the Schr\"odinger equation 
is given by adiabatic states of $\hat{H}_0(t)$.
Then, it would be an interesting problem to apply this separation
to general nonequilibrium processes.

In STA, a cost of the time evolution was studied in~\cite{DR3, SS, CSHS, ZCCP}.
A trade-off relation between time, energy fluctuation and state distance is 
known as the quantum speed limit~\cite{MT} and was discussed 
in the context of STA in~\cite{CD}.
The applications of STA to thermodynamic systems were studied 
in several works~\cite{DWLHG, dCGP, XG, BJdC}.
A universal trade-off relation was derived from 
work fluctuation relations in~\cite{FZCKUC}.

In this paper, we study properties of the nonequilibrium entropy production
that are applicable to general nonequilibrium processes.
In thermally-isolated systems, 
the entropy is directly related to the work average 
and the present result is essentially equivalent
to the result in~\cite{FZCKUC}. 
However, the entropy is represented by the Kullback--Leibler (KL) divergence, 
which leads us naturally to 
the information-geometric interpretation of the nonequilibrium process.
Establishing this novel picture is the main aim of the present work.

The organization of this paper is as follows.
In section~\ref{sec:sta}, we discuss how a given Hamiltonian is 
separated into two parts.
Then, the method is applied to the nonequilibrium entropy production 
in section~\ref{sec:entropy}. 
In section~\ref{sec:bound}, we discuss lower bounds of the entropy
by using the improved Jensen inequalities and derive a trade-off relation.
The last section~\ref{sec:conclusion} is devoted to conclusions.

%%%%%%%%%%%%%%%%%%%%%%%%%%%%%%%%%%%%%%%%%%%%%%%%%%%%%%%%%%%%%%%%%%%%%%%%%%%
\section{Shortcuts to adiabaticity for general dynamical systems}
\label{sec:sta}

%%%%%%%%%%%%%%%%%%%%%%%%%%%%%%%%%%%%%%%%%%%%%%%%%%%%%%%%%%%%%%%%%%%%%%%%%%%
\subsection{General formula}

We start from reviewing the method of STA, 
somewhat in a different way from the standard prescription~\cite{DR1, Berry}, 
with the emphasis on its applicability to general dynamical systems.
For a given time-dependent Hamiltonian $\hat{H}(t)$
and an initial state $|\psi(0)\rangle$, the time evolution of the state, 
$|\psi(t)\rangle$, satisfies the Schr\"odinger equation
\be
 i\frac{\partial}{\partial t}|\psi(t)\rangle = \hat{H}(t)|\psi(t)\rangle,
\ee
where we put $\hbar=1$.
When we start the time evolution from an eigenstate of the initial 
Hamiltonian $|n\rangle$ satisfying 
$\hat{H}(0)|n\rangle=\epsilon_n(0)|n\rangle$, 
the state is written as 
\be
 |\psi_n(t)\rangle = \hat{U}(t)|n\rangle, 
\ee
where $\hat{U}(t)$ is the time evolution operator.
Generally, the state vector is defined on a Hilbert space 
and the total number of indices 
is equal to the dimension of the space.
The eigenstates $\{|n\rangle\}$ satisfy the orthonormal relation 
$\langle m|n\rangle=\delta_{m,n}$ and 
the completeness relation $\sum_n |n\rangle\langle n|=1$.

We write the Hamiltonian using the basis $\{|\psi_n(t)\rangle\}$.
The Hamiltonian is separated into 
the diagonal and offdiagonal parts: 
$\hat{H}(t)=\hat{H}_0(t)+\hat{H}_1(t)$.
They are written respectively as 
\be
 && \hat{H}_0(t)=\sum_n 
 |\psi_n(t)\rangle\langle\psi_n(t)|\hat{H}(t)|\psi_n(t)\rangle
 \langle \psi_n(t)|, \label{H0} \\
 && \hat{H}_1(t)
 = \sum_{m\ne n} 
 |\psi_m(t)\rangle\langle \psi_m(t)|\hat{H}(t)|\psi_n(t)\rangle\langle \psi_n(t)|.
\ee
This separation indicates that the state $|\psi(t)\rangle$ 
is given by the eigenstates of $\hat{H}_0(t)$ with the eigenvalue 
$\epsilon_n(t)=\langle \psi_n(t)|\hat{H}(t)|\psi_n(t)\rangle$.
The most general form of the state is 
\be
 |\psi(t)\rangle = \sum _n c_n |\psi_n(t)\rangle,
\ee
where $c_n$ is a time-independent constant.
$\hat{H}_1(t)$ is called the counterdiabatic term and 
is rewritten by using the Schr\"odinger equation as 
\be
 \hat{H}_1(t)= i\sum_{m\ne n} 
 |\psi_m(t)\rangle\langle\psi_m(t)|\dot{\psi}_n(t)\rangle
 \langle \psi_n(t)|, \label{dot}
\ee
where the dot symbol denotes the time derivative.
The counterdiabatic term prevents nonadiabatic transitions between 
instantaneous eigenstates of $\hat{H}_0(t)$.

We note that the eigenstate of $\hat{H}_0(t)$ does not necessarily satisfy 
the Schr\"odinger equation.
Following the definition of the adiabatic state, 
we need to add an appropriate phase as 
\be
 |\psi_n(t)\rangle =\exp\left(-i\int_0^t \diff t'\,\epsilon_n(t')
 -\int_0^t \diff t'\,\langle n(t')|\dot{n}(t')\rangle\right)|n(t)\rangle, 
\ee
where $|n(t)\rangle$ is an eigenstate of $\hat{H}_0(t)$.
Using the eigenstate set $\{|n(t)\rangle\}$, we can write the Hamiltonians
(\ref{H0}) and (\ref{dot}) 
in the same form with the replacement $|\psi_n(t)\rangle\to |n(t)\rangle$.

Thus, the problem of solving the Schr\"odinger equation 
for a given Hamiltonian $\hat{H}(t)$ 
reduces to finding the proper separation $\hat{H}(t)=\hat{H}_0(t)+\hat{H}_1(t)$.
In the engineering problem, 
we consider $\hat{H}_0(t)$ as the original Hamiltonian
and the additional counterdiabatic term is introduced 
to prevent the nonadiabatic transitions.
However, this procedure is problematic in most cases
since the counterdiabatic term 
generally takes a complicated form and is hard to manipulate~\cite{OT1}.
Otherwise, we can consider the inverse engineering to keep
the original form of the Hamiltonian~\cite{CRSCGM}.
Here we set up the problem by defining the total Hamiltonian 
so that the method is applicable to any dynamical systems.
Although the separation of the Hamiltonian is possible in any systems,  
it is generally a difficult problem to find the proper separation.

The time dependence of the Hamiltonian appears through 
parameters in the Hamiltonian. 
Since the state is given by the instantaneous eigenstates 
of $\hat{H}_0$, we consider that  
the eigenstates and eigenvalues depend on parameters $\lambda(t)$ 
as $|n(\lambda(t))\rangle$ and $\epsilon_n(\lambda(t))$ respectively. 
On the other hand, the time derivative appears in $\hat{H}_1$ 
which means that the counterdiabatic term is written 
as $\hat{H}_1=\dot{\lambda}(t)\hat{\xi}(\lambda(t))$ where 
\be
 \hat{\xi}(\lambda)= i\sum_{m\ne n} 
 |m(\lambda)\rangle\langle m(\lambda)|\partial_\lambda n(\lambda)\rangle
 \langle n(\lambda)|.
\ee
It was discussed in \cite{OT2} that, for classical systems, 
$\xi$ is represented by the $\lambda$-derivative of a generalized action.
The action is introduced by using the Hamilton--Jacobi theory and 
reduces to the adiabatic invariant in a special case, which imply that
the counterdiabatic term characterizes the dynamics.
Although we discuss quantum systems in this paper, 
the following discussions are applicable to classical systems as well.

%%%%%%%%%%%%%%%%%%%%%%%%%%%%%%%%%%%%%%%%%%%%%%%%%%%%%%%%%%%%%%%%%%%%%%%%%%%
\subsection{Quantum quench}

It is worth mentioning that 
the method of STA is applicable even when 
a prepared initial state is driven by a time-independent total Hamiltonian.
Using the time-dependent basis, we can introduce a time-dependent 
$\hat{H}_0(t)$ and $\hat{H}_1(t)$ to write $\hat{H}=\hat{H}_0(t)+\hat{H}_1(t)$.
We note that the separation is useful only when the initial state is
not in the eigenstate of the Hamiltonian.
Such a situation is realized in the problem of quantum quench 
where we consider the state evolution 
under a sudden change of the Hamiltonian~\cite{quench}.

First, we prepare the state as an eigenstate of the Hamiltonian $\hat{H}^{(0)}$.
The eigenstate $|n\rangle$ satisfies the eigenvalue equation 
\be
 \hat{H}^{(0)}|n\rangle = \epsilon_n^{(0)}|n\rangle.
\ee
Then, at $t=0$, we start the state evolution 
by a different Hamiltonian $\hat{H}$.
The state is given by $|\psi_n(t)\rangle =\e^{-i\hat{H}t}|n\rangle$ 
where we set the initial condition $|\psi_n(0)\rangle=|n\rangle$.

As we explained in the general formulation, the Hamiltonian is 
separated into two parts by using the basis $|\psi_n(t)\rangle$.
Using the fact that the total Hamiltonian is time-independent at $t>0$, 
we can write $\hat{H}=\hat{H}_0(0)+\hat{H}_1(0)$ where 
$\hat{H}_0(0)$ is the diagonal part with respect to the basis $\{|n\rangle\}$
and satisfies $[\hat{H}^{(0)},\hat{H}_0(0)]=0$.
$\hat{H}_1(0)$ is the offdiagonal part and is defined by $\hat{H}-\hat{H}_0(0)$.
For the time-evolved state, 
$\hat{H}_0(t)$ and $\hat{H}_1(t)$ are written respectively as 
\be
 && \hat{H}_0(t) = \e^{-i\hat{H}t}\hat{H}_0(0)\e^{i\hat{H}t}
 = \sum_n\epsilon_n |\psi_n(t)\rangle\langle \psi_n(t)|, \\
 && \hat{H}_1(t) = \e^{-i\hat{H}t}\hat{H}_1(0)\e^{i\hat{H}t}, 
\ee
where $\epsilon_n=\langle \psi_n(t)|\hat{H}|\psi_n(t)\rangle$ 
is time independent.
The problem of quantum quench is reduced to solving 
the eigenvalue equation if we know the form of $\hat{H}_0(t)$.
Of course, it is still a difficult problem in general.

%%%%%%%%%%%%%%%%%%%%%%%%%%%%%%%%%%%%%%%%%%%%%%%%%%%%%%%%%%%%%%%%%%%%%%%%%%%
\subsection{Example}

The simplest example is the system where 
the dimension of the Hilbert space is equal to two.
The general Hamiltonian is written by using the Pauli-operator vector 
$\hat{\bm{\sigma}}$ as 
\be
 \hat{H}(t)=\frac{1}{2}\bm{h}(t)\cdot\hat{\bm{\sigma}}
 =\frac{1}{2}
 \left(\bm{h}_0(t)+\frac{\bm{h}_0(t)\times\dot{\bm{h}}_0(t)}{\bm{h}_0^2(t)}
 \right)\cdot\hat{\bm{\sigma}}. \label{two}
\ee
The second equality denotes the separation of $\hat{H}_0(t)$ and $\hat{H}_1(t)$.
For a given vector function $\bm{h}(t)$, 
we need to find $\bm{h}_0(t)$.
Although the explicit general formula 
to write $\bm{h}_0(t)$ in terms of $\bm{h}(t)$ is not known,
it is clear from the above general discussion that
such a function $\bm{h}_0(t)$ can be obtained in principle. 

An example where the total Hamiltonian is time-independent 
was treated in~\cite{Takahashi1}.
We exploit that example to see below how the method works 
when it is applied to the problem of quantum quench.

We consider $\bm{h}=(0,0,h)$ where $h$ is a real constant.
In this case, the Schr\"odinger equation is easily solved by 
the standard textbook method.
The general form of the state is given by 
\be
 |\psi(t)\rangle 
 = c_+\e^{-\frac{iht}{2}}|+\rangle
 +c_-\e^{\frac{iht}{2}}|-\rangle, \label{stat}
\ee
where $|\pm\rangle$ are eigenstates of $\hat{\sigma}^z$ satisfying 
$\hat{\sigma}^z|\pm\rangle = \pm |\pm\rangle$, and 
$c_{\pm}$ are complex constant values determined from the initial condition.
If we start the time evolution from one of the eigenstates, 
the state remains in the same eigenstate throughout the time evolution.

We analyze the same system by using STA.
Using the formula in (\ref{two}), 
we can find the most general form of $\bm{h}_0(t)$ as 
\be
 \bm{h}_0(t)=h\cos\theta_0\bmat{c}
 \sin\theta_0\cos(ht+\varphi_0) \\
 \sin\theta_0\sin(ht+\varphi_0) \\
 \cos\theta_0 \emat, 
\ee
where $\theta_0$ and $\varphi_0$ are real constants.
Each part of the Hamiltonian is given respectively as 
\be
 && \hat{H}_0(t)=\frac{1}{2}h\cos\theta_0\bmat{c}
 \sin\theta_0\cos(ht+\varphi_0) \\
 \sin\theta_0\sin(ht+\varphi_0) \\
 \cos\theta_0 \emat\cdot\hat{\bm{\sigma}}, \\
 && \hat{H}_1(t)=-\frac{1}{2}h\sin\theta_0\bmat{c}
 \cos\theta_0\cos(ht+\varphi_0) \\
 \cos\theta_0\sin(ht+\varphi_0) \\
 -\sin\theta_0 \emat\cdot\hat{\bm{\sigma}}.
\ee
In this example, we see that the time-dependent parameter 
is given by $\lambda(t)=ht$.
The corresponding state is given by a linear combination of 
the adiabatic states of $\hat{H}_0(t)$.
We obtain  
\be
 |\psi(t)\rangle 
 = \tilde{c}_+\e^{-\frac{iht}{2}}
 \bmat{c} \cos\frac{\theta_0}{2} \\ 
 \e^{i(ht+\varphi_0)}\sin\frac{\theta_0}{2} \emat
 + \tilde{c}_-\e^{\frac{iht}{2}}
 \bmat{c} -\e^{-i(ht+\varphi_0)}\sin\frac{\theta_0}{2} \\
 \cos\frac{\theta_0}{2} \emat,
\ee
where the vector representation 
$\bmat{cc}a & b \emat^{\rm T}=a|+\rangle+b|-\rangle$ is used  
and $\tilde{c}_\pm$ are complex constant values.
This state is equivalent to (\ref{stat}) but 
the separation of the vector has a different meaning.
If we start the time evolution from one of the eigenstates 
for $\hat{H}_0(0)$ (one of two vectors at $t=0$ in the above equation), 
the state remains in the same eigenstate of $\hat{H}_0(t)$ 
(one of two vectors at $t$ in the above equation) 
throughout the time evolution.
This picture holds even when the initial state is not 
in the eigenstate of the initial Hamiltonian.
We note that the eigenstate is time dependent in this case.
This result will be a useful tool to understand the quench dynamics.

The important conclusion in this section is that 
the separation of the Hamiltonian is very general and 
is applied for arbitrary choices of the Hamiltonian
as we see in the above derivation.  
This means that the general dynamics is characterized by STA.
As a possible application, we consider the nonequilibrium 
entropy production in the following.

%%%%%%%%%%%%%%%%%%%%%%%%%%%%%%%%%%%%%%%%%%%%%%%%%%%%%%%%%%%%%%%%%%%%%%%%%%%
%%%%%%%%%%%%%%%%%%%%%%%%%%%%%%%%%%%%%%%%%%%%%%%%%%%%%%%%%%%%%%%%%%%%%%%%%%%
\section{Nonequilibrium entropy production}
\label{sec:entropy}

%%%%%%%%%%%%%%%%%%%%%%%%%%%%%%%%%%%%%%%%%%%%%%%%%%%%%%%%%%%%%%%%%%%%%%%%%%%
\subsection{Entropy production and Pythagorean theorem}

To characterize nonequilibrium states, we use 
the work done on the system as one of the measure.
We prepare the state in contact with a bath and 
the initial state is in equilibrium.
Then, the system is thermally isolated from the outside and 
is evolved under a control by the external agent.
The work is obtained by the two-time measurement scheme.

The initial state is assumed to be distributed 
according to the Boltzmann distribution 
$p_n(0)=\e^{-\beta \epsilon_n(0)}/Z_0$ where $Z_0=\sum_n \e^{-\beta \epsilon_n(0)}$
and $\beta$ is the inverse temperature.
The time evolution of the system is described by 
the time-dependent Hamiltonian $\hat{H}(t)$ and the work is 
defined by the energy difference between the initial and final states. 
Since the initial state is distributed randomly, we can define
the work distribution function 
\be
 P(W,t) &=& \sum_n p_n(0)\langle \psi_n(t)|\delta
 \left(W-(\hat{H}(t)-\epsilon_n(0))\right)|\psi_n(t)\rangle \no\\
 &=& \sum_n p_n(0)\langle n(t)|\delta
 \left(W-(\hat{H}(t)-\epsilon_n(0))\right)|n(t)\rangle. \label{wdf}
\ee
The main question here is whether we can find any useful information 
on this work distribution 
by using the separation of the Hamiltonian 
$\hat{H}(t)=\hat{H}_0(t)+\hat{H}_1(t)$.

We are mainly interested in the work average $[W]_t=\int \diff W\, P(W,t)W$.
In \cite{FZCKUC}, it was shown that the average is given by 
\be
 [W]_t 
 = \sum_n p_n(0)\langle n(t)|(\hat{H}(t)-\epsilon_n(0))|n(t)\rangle
 = \sum_n p_n(0)(\epsilon_n(t)-\epsilon_n(0)), \label{W}
\ee
with the use of the relation $\langle n(t)|\hat{H}_1(t)|n(t)\rangle=0$.
This equation shows that 
the counterdiabatic term $\hat{H}_1(t)$ does not contribute to the average.
It can also be shown that the squared average $[W^2]_t$ is 
separated into two parts as
\be
 [W^2]_t 
 = \sum_n p_n(0)(\epsilon_n(t)-\epsilon_n(0))^2
 + \sum_n p_n(0)\langle n(t)|\hat{H}_1^2(t)|n(t)\rangle, \label{W2}
\ee
and the second term has a geometric meaning as we discuss below.

Using the averaged work, we define the nonequilibrium entropy production 
\be
 \Sigma(t)=\beta [W]_t-\beta(F_t-F_0),
\ee
where $F_t$ is the free energy for the Hamiltonian  $\hat{H}(t)$ 
and is defined as $-\beta F_t=\ln Z_t$.
$Z_t=\Tr \e^{-\beta \hat{H}(t)}$ is the partition function 
defined at each $t$. 
We note that the final state is not necessarily in equilibrium.

$\Sigma(t)$ is a nonnegative quantity.
This property is understood from the relation that $\Sigma(t)$ is written 
by the KL divergence of two density operators:  
\be
 \Sigma(t)= D_{\rm KL}(\hat{\rho}(0\!\to\!t)||\hat{\rho}(t)).
\ee
We use (\ref{W}) to derive this equation.
The KL divergence is defined as $D_{\rm KL}(\hat{P}||\hat{Q})
=\Tr \hat{P}\ln\hat{P}-\Tr \hat{P}\ln\hat{Q}$ and the density operators 
are defined as
\be
 && \hat{\rho}(0\!\to\!t) 
 = \frac{1}{Z_0}\hat{U}(t)\e^{-\beta\hat{H}(0)}\hat{U}^\dag(t) 
 = \frac{1}{Z_0}\exp\left(-\beta\hat{U}(t)\hat{H}(0)\hat{U}^\dag(t)\right), 
 \label{rho0t} \\
 && \hat{\rho}(t) = \frac{1}{Z_t}\e^{-\beta\hat{H}(t)}.
\ee
$\hat{\rho}(0\!\to\!t)=\hat{U}(t)\hat{\rho}(0)\hat{U}^\dag(t)$ 
is the time-evolved state of the initial distribution $\hat{\rho}(0)$ and 
represents the actual distribution of states at each $t$.
On the other hand, $\hat{\rho}(t)$ represents the distribution
for the canonical equilibrium states of the Hamiltonian $\hat{H}(t)$.
Generally, the evolved state is not in equilibrium and 
these operators are not equal with each other.
The above relation shows that the entropy production 
represents how far the nonequilibrium state is from the equilibrium one 
and is written by the divergence of two distributions.

As we see in (\ref{rho0t}), 
$\hat{\rho}(0\!\to\!t)$ is characterized by the effective Hamiltonian 
$\hat{H}(0\!\to\!t)=\hat{U}(t)\hat{H}(0)\hat{U}^\dag (t)$.
Its spectral decomposition is given by 
\be
 \hat{H}(0\!\to\!t)=\sum_n \epsilon_n(0)|n(t)\rangle\langle n(t)|, 
\ee
which has a similar form to $\hat{H}_0(t)$.
The eigenstates are time dependent but the eigenvalues are not.
This Hamiltonian satisfies the equation 
for the Lewis--Riesenfeld invariant~\cite{LR}: 
\be
 i\frac{\partial \hat{H}(0\!\to\!t)}{\partial t}
 =[\hat{H}(t), \hat{H}(0\!\to\!t)]
 =[\hat{H}_1(t), \hat{H}(0\!\to\!t)].
\ee
We note that $\hat{H}_0(t)$ and $\hat{H}(0\!\to\!t)$ commutes with each other.
This equation was studied systematically in \cite{OT1}  
and the Lax form for classical nonlinear integrable systems
is shown to be useful to find a pair $(\hat{H}(0\!\to\!t),\hat{H}_1(t))$.
The relation $\hat{H}_0(t)=\hat{H}(0\!\to\!t)$ holds when the energy eigenvalues 
of $\hat{H}_0(t)$ are independent of $t$.
In this special case, the entropy is given by the divergence 
between two canonical distributions 
$\hat{\rho}_0(t) = \e^{-\beta\hat{H}_0(t)}/Z_t^{(0)}$ 
and $\hat{\rho}(t)$ as 
\be
 \Sigma(t)=D_{\rm KL}(\hat{\rho}(0\!\to\!t)||\hat{\rho}(t))
 = D_{\rm KL}(\hat{\rho}_0(t)||\hat{\rho}(t))
 = \beta (F_t^{(0)}-F_t),
\ee
where $-\beta F_t^{(0)}=\ln Z_t^{(0)}=\ln\Tr \e^{-\beta\hat{H}_0(t)}$.
The entropy is given by the free energy difference.
Here we use again $\langle n(t)|\hat{H}_1(t)|n(t)\rangle=0$.

In the general case $\hat{H}_0(t)\ne \hat{H}(0\!\to\!t)$, 
by using the separation of the Hamiltonian 
$\hat{H}(t)=\hat{H}_0(t)+\hat{H}_1(t)$, 
we can easily show the entropy production
is separated into two parts: 
\be
 \Sigma(t) = D_{\rm KL}(\hat{\rho}(0\!\to\!t)||\hat{\rho}(t))
 =\Sigma_0(t)+\Sigma_1(t).  \label{pythagorean}
\ee
Each term is written by using the KL divergence: 
\be
 \Sigma_0(t) &=& D_{\rm KL}(\hat{\rho}(0\!\to\!t)||\hat{\rho}_0(t)) \no\\
 &=& \frac{1}{Z_0}\sum_n \e^{-\beta\epsilon_n(0)}\beta(\epsilon_n(t)-\epsilon_n(0))
 +\beta (F_0-F_t^{(0)}),
\ee
\be
 \Sigma_1(t) = D_{\rm KL}(\hat{\rho}_0(t)||\hat{\rho}(t))
 = \beta (F_t^{(0)}-F_t).
\ee
$\Sigma_0(t)$ represents the KL divergence 
between the canonical distributions of 
$\hat{H}(0\!\to\!t)$ and $\hat{H}_0(t)$ and is expressed by 
the spectrum distance between $\{\epsilon_n(0)\}$ and $\{\epsilon_n(t)\}$.
It is independent of $\hat{H}_1(t)$. 
On the other hand, $\Sigma_1(t)$ represents a distance due to 
the counterdiabatic term since it goes to zero when $\hat{H}_1(t)=0$.
Thus, the entropy production is separated into two parts and each 
part plays a different role.

We note that the difference between $\Sigma(t)$ and $\Sigma_1(t)$ 
has been studied in some works.
In~\cite{dCGP}, the difference was studied 
for a process in an Otto cycle and the result 
was plotted for the harmonic-oscillator Hamiltonian.
In~\cite{DPZ}, $\Sigma_1(t)$ was defined 
in a process of the projective measurement 
to derive the inequality $\Sigma(t)\ge \Sigma_1(t)$.
Our result is derived as an equality and
is applicable to general systems.

%%%%%%%%%%%%%%%%%%%%%%%%%%%%%%%%%%%%%%%%%%%%%%%%%%%%%%%%%%%%%%%%%%%%%%%%%%%
\subsection{Information-geometric interpretation}

It is well known that the KL divergence is 
a generalization of a squared distance measure.
This means that (\ref{pythagorean}) represents the Pythagorean theorem 
and has a geometric meaning. 
The theorem has been closely discussed 
in the field of information geometry~\cite{Amari}.
In the following, we interpret the result (\ref{pythagorean}) 
to refine the method of STA  
from the aspect of the information geometry~\cite{Amari}.

The Hamiltonian is generally written as 
\be
 -\beta\hat{H}(\theta)=\sum_i \theta_i\hat{X}_i.
\ee
Although the following discussions hold for classical systems as well, 
we use general finite-dimensional quantum systems for the description.
Then, 
$\{\hat{X}_i\}$ represents a set of independent operators and 
the number of operators is determined by specifying the Hilbert space.
The coefficient $\theta=(\theta_1,\theta_2,\dots)$ plays the role
of coordinates.
The coordinate system is used to specify the probability distribution.
In the present study we treat the canonical distribution 
\be
 \hat{\rho}(\theta)=\exp\left(-\beta\hat{H}(\theta)-\psi(\theta)\right).
 \label{rhotheta}
\ee
The normalization function $\psi(\theta)$ defined as 
$\psi(\theta)=\ln\Tr \e^{-\beta\hat{H}(\theta)}$ 
is a convex function and represents the free energy $-\beta F$ in physics.

For a coordinate system $\theta$ where 
a convex function $\psi(\theta)$ is defined, 
we can introduce the dual coordinate system $\theta^*$ and
the dual convex function $\psi^*(\theta^*)$ 
by using the Legendre transformation.
The dual coordinate is defined by 
\be
 \theta^*_i = \frac{\partial}{\partial\theta_i}\psi(\theta), 
\ee
and the corresponding convex function is 
\be
 \psi^*(\theta^*)=\theta\cdot\theta^*-\psi(\theta), 
\ee
where we use the abbreviation 
$\theta\cdot\theta^*=\sum_i\theta_i^{\phantom{*}}\theta^*_i$.
In the canonical distribution, the dual coordinate is written as
the canonical average of operators: 
\be
 \theta^*_i = \langle\hat{X}_i\rangle_{H}
 =\Tr \hat{X}_i\exp\left(-\beta\hat{H}(\theta)-\psi(\theta)\right).
\ee
The state in the canonical distribution
can also be uniquely specified by $\theta^*$ instead of using $\theta$.

In a coordinate system where a convex function $\psi(\theta)$ is defined, 
the Bregman divergence is introduced as a distance measure
\be
 D_\psi(\theta||\theta')
 =\psi(\theta)-\psi(\theta')-\theta'^{*}\cdot(\theta-\theta')
 =\psi(\theta)+\psi^*(\theta'^*)-\theta\cdot\theta'^{*}.
\ee
The last expression shows that the dual divergence can also be defined 
as $D_{\psi}^*(\theta^*||\theta'^*)=D_\psi(\theta'||\theta)$.
We note that the divergence is not symmetric in general:   
$D_\psi(\theta||\theta')\ne D_\psi(\theta'||\theta)$.
However, by considering an infinitesimal distance, 
it has a symmetric form, which defines 
the Riemannian metric in the manifold parametrized by 
the coordinate $\theta$.
In the present case where the probability distribution is given by 
the canonical distribution,
the Bregman divergence is equivalent to the KL divergence: 
\be
 D_\psi(\theta||\theta')=D_{\psi}^*(\theta'^*||\theta^*)
 =D_{\rm KL}(\theta'||\theta).
\ee

Now we discuss the geometric meaning of the Pythagorean theorem.
Since we are interested in the KL divergence, 
we treat the dual divergence 
$D^*_\psi(\theta^*||\theta'^*)=D_{\rm KL}(\theta||\theta')$.
For given three points $\theta_P$, $\theta_Q$ and $\theta_R$, 
We consider the condition such that 
the dual Pythagorean theorem holds: 
\be
 D_{\psi}^*(P||R)=D_{\psi}^*(P||Q)+D_{\psi}^*(Q||R).
\ee
A simple calculation gives 
\be
 (\theta^*_P-\theta^*_Q)\cdot (\theta_Q-\theta_R)=0. \label{perp}
\ee
The affine coordinate system $\theta$ introduced in (\ref{rhotheta}) represents   
a dually-flat manifold.
The geodesic is parametrized as a straight line connecting two points,
Q and R, as $\theta_{QR}(\tau)=\tau\theta_Q+(1-\tau)\theta_R$
where $\tau$ parametrizes the straight line and takes 
between 0 and 1.
Then, $\theta_Q-\theta_R$ represents the tangent vector.
In the same way the dual geodesic line is written by 
using the dual coordinate.
Equation (\ref{perp}) means that the dual geodesic 
$\theta_P^*-\theta_Q^*$ is 
perpendicular to the geodesic $\theta_Q-\theta_R$.

We apply the general argument to the present problem.
Three points, P, Q and R, are represented by 
the canonical distributions of 
$\hat{H}(0\!\to\!t)$, $\hat{H}_0(t)$ and $\hat{H}(t)$ respectively.
Then, the counterdiabatic term is written as 
\be
 -\beta\hat{H}_1(t)=\left(\theta_R(t)-\theta_Q(t)\right)\cdot\hat{X}.
\ee
To identify $\theta_P^*-\theta_Q^*$, we note that 
the dual geodesic is represented 
by the canonical average of operators. 
The point P is the average in terms of 
$\hat{H}(0\!\to\!t)$ and Q is in terms of $\hat{H}_0(t)$.
These Hamiltonians are diagonalized by the basis $\{|n(t)\rangle\}$.
Then, the dual geodesic is defined on a submanifold where 
the Hamiltonian is diagonalized by the same basis.
Using the relation $\langle n(t)|\hat{H}_1(t)|n(t)\rangle=0$,
we conclude that the perpendicular condition is represented as 
\be
 0 &=& -\beta\langle\hat{H}_1(t)\rangle_{\hat{H}(0\to t)}
 +\beta\langle\hat{H}_1(t)\rangle_{\hat{H}_0(t)} \no\\
 &=&\left(\theta_R(t)-\theta_Q(t)\right)\cdot\left(
 \langle\hat{X}\rangle_{\hat{H}(0\to t)}- \langle\hat{X}\rangle_{\hat{H}_0(t)}
 \right) \no\\
 &=& \left(\theta_R(t)-\theta_Q(t)\right)\cdot
 \left(\theta_P^*(t)-\theta_Q^*(t)\right).
\ee

The dual geodesic connecting P and Q is interpreted 
the dual projection of P to a flat submanifold.
The flat submanifold includes points on the geodesic QR, 
and is parametrized by the coordinate $\theta$.
Each point is represented by the canonical distribution of the Hamiltonian 
$\hat{H}_e(t)=\hat{H}_0(t)+\theta_e(t)\cdot\hat{X}$.
The coordinate $\theta_e(t)$ satisfies the perpendicular condition 
\be
 \langle n(t)|\theta_e(t)\cdot\hat{X}|n(t)\rangle = 0.
\ee
The point Q represents the nearest point of P in the ``e-flat'' submanifold.
In the same way, we can define the ``m-flat'' submanifold 
including the dual geodesic PQ, 
which is parametrized by the dual coordinate.
Then, the geodesic RQ represents the projection of the point R
to the submanifold.
This property is known as the projection theorem
in the information geometry.

%%%%%%%%%%%%%%
\begin{figure}[t]
\begin{center}
\includegraphics[width=0.5\columnwidth]{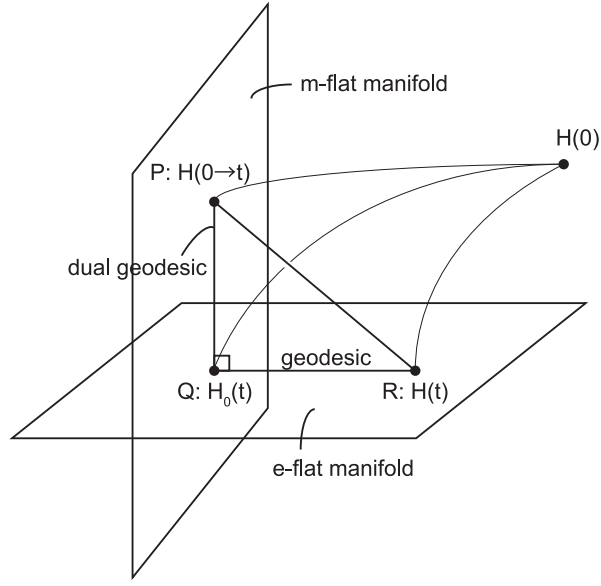}
\caption{Information-geometric interpretation of 
the nonequilibrium entropy production.
The Hamiltonian is evolved from $\hat{H}(0)$ to three different forms 
$\hat{H}(0\!\to\!t)=\hat{U}(t)\hat{H}(0)\hat{U}^\dag(t)$, 
$\hat{H}_0(t)$ and $\hat{H}(t)=\hat{H}_0(t)+\hat{H}_1(t)$, and 
their canonical distributions are  
denoted by P, Q and R respectively.
Then, the points make a right triangle and 
the entropy production satisfies the Pythagorean theorem  
$D_{\rm KL}(P||R)=D_{\rm KL}(P||Q)+D_{\rm KL}(Q||R)$.
The dual geodesic PQ is perpendicular to the geodesic QR.}
\label{fig:manifold}
\end{center}
\end{figure}
%%%%%%%%%%%%

We summarize the information-geometric interpretation 
of the Pythagorean theorem in Figure~\ref{fig:manifold}.
This interpretation shows that 
the Hamiltonian $\hat{H}_0(t)$ plays an important role 
for the difference between $\hat{H}(0\!\to\!t)$ and $\hat{H}(t)$.
For a given flat manifold including R, the point Q is uniquely determined
by the dual projection of P to the manifold.
Of course, it is a difficult problem to find the proper manifold and 
this interpretation is of no use in general 
to find $\hat{H}_0(t)$ for a given $\hat{H}(t)$.
Nevertheless, we expect that 
this new picture will be a guiding principle to design the system.

%%%%%%%%%%%%%%%%%%%%%%%%%%%%%%%%%%%%%%%%%%%%%%%%%%%%%%%%%%%%%%%%%%%%%%%%%%%
\section{Lower bounds of entropy production}
\label{sec:bound}

%%%%%%%%%%%%%%%%%%%%%%%%%%%%%%%%%%%%%%%%%%%%%%%%%%%%%%%%%%%%%%%%%%%%%%%%%%%
\subsection{Improved Jensen inequality}

In the previous section, we studied that the nonequilibrium entropy 
is separated into two parts according to their roles 
and is represented by the KL divergence. 
The KL divergence represents a degree of separation 
of two points and the metric is obtained from an infinitesimal separation.

It is an interesting problem to study the lower bound of the entropy production
since it characterizes the efficiency of the system control.
The counterdiabatic term has a geometric meaning and is related to 
the Fubini--Study distance~\cite{CHKO, Takahashi2}.
In the setting we are studying in this paper, 
we consider a time evolution from $\hat{\rho}(0)$ to 
$\hat{\rho}(0\!\to\!t)=\hat{U}(t)\hat{\rho}(0)\hat{U}^\dag(t)$.
The length between two state is defined, according to \cite{FZCKUC}, by 
\be
 \ell(\hat{\rho}(0), \hat{\rho}(0\!\to\!t)) 
 &=& \int_0^t\diff t'\,\sqrt{\sum_n p_n(0)\langle \dot{n}(t')|\hat{H}_1^2(t')
 |\dot{n}(t')\rangle} \no\\ 
 &=& 
 \int_0^t\diff t'\,\sqrt{\sum_n p_n(0)\langle \dot{n}(t')|
 \left(1-|n(t')\rangle\langle n(t')|\right)
 |\dot{n}(t')\rangle}. \label{ell}
\ee 
This is considered to be a natural length since it satisfies 
$\ell(\hat{\rho}(0), \hat{\rho}(0\!\to\!t)) \ge {\cal L}(\hat{\rho}(0), \hat{\rho}(0\!\to\!t))$ 
where ${\cal L}$ represents the Bures distance: 
\be
 {\cal L}(\hat{\rho}_1,\hat{\rho}_{2}) ={\rm arccos}\,
 \Tr \sqrt{\sqrt{\hat{\rho}_1}\hat{\rho}_2\sqrt{\hat{\rho}_1}}.
\ee
We note that the integrand in $\ell$ appears in 
the squared average of the work as in (\ref{W2}).
In the nonequilibrium entropy production,  
we treat the divergence between $\hat{\rho}(0\!\to\!t)$ and $\hat{\rho}(t)$
and it is expected to be bounded from below by an appropriate distance.
From a mathematical point of view, 
the KL divergence is bounded below by the Bures distance.
It was shown in \cite{DL1, DL2, AE} that 
\be
 D_{\rm KL}(\hat{\rho}_1||\hat{\rho}_{2})\ge \frac{8}{\pi^2}
 {\cal L}^2(\hat{\rho}_1,\hat{\rho}_{2}). \label{L2}
\ee
When this relation is applied to the entropy production 
$\Sigma(t)=D_{\rm KL}(\hat{\rho}(0\!\to\!t)||\hat{\rho}(t))$, 
we can find a lower limit.
However, this relation holds at each $t$ and does not suitable to 
characterize the efficiency of the time evolution.
In this section, we study a different lower bound of the nonequilibrium entropy 
by using the improved Jensen inequality and 
apply it to derive a trade-off relation for the time evolution.

The property $\Sigma(t)\ge 0$ can be 
shown by using the Jensen inequality 
\be
 \e^{-\beta[W]_t}\le [\e^{-\beta W}]_t =\e^{-\beta(F_t-F_0)}, \label{jensen0}
\ee
where the equality represents the Jarzynski formula obtained 
from the definition of the work distribution (\ref{wdf}).
To find a nontrivial lower bound of $\Sigma(t)$, 
we use the improved Jensen inequality 
derived in~\cite{Decoster}.
Using the formula in \ref{sec:jensen}, we obtain 
\be
  \e^{-\beta[W]_t}\le \frac{[\e^{-\beta W}]_t}{1+\frac{\beta^2}{2}
 C(t)[(W-[W]_t)^2]_t},
\ee
where $C(t)$ is a positive function and is written as  
\be
 C(t)=\exp\left(
 -\frac{\beta}{3}\frac{[(W-[W]_t)^3]_t}{[(W-[W]_t)^2]_t}\right).
\ee
This gives a tighter bound of $\Sigma(t)$ than (\ref{jensen0}).
We can write 
\be
 \Sigma(t)
 \ge \ln \left(1+\frac{\beta^2}{2}C(t)
 [(W-[W]_t)^2]_t\right). \label{bound}
\ee
The bound is written by the second variance of $W$.
Using (\ref{W}) and (\ref{W2}), we have 
\be
 [(W-[W]_t)^2]_t\ge 
 \sum_n p_n(0)\langle n(t)|\hat{H}_1^2(t)|n(t)\rangle. \label{W2c}
\ee
The form of the right-hand side appears in (\ref{ell}) and we can write 
\be
 \int_0^t \diff t'\,\sqrt{\e^{\Sigma(t')}-1} 
 &\ge& 
 \sqrt{\frac{\beta^2}{2}C_{\rm min}}\,\ell(\hat{\rho}(0),\hat{\rho}(0\!\to\!t)) \no\\
 &\ge&
 \sqrt{\frac{\beta^2}{2}C_{\rm min}}\,{\cal L}(\hat{\rho}(0),\hat{\rho}(0\!\to\!t)),
\ee
where $C_{\rm min}$ represents the minimum value of $C(t)$.
This inequality represents a trade-off relation.
The left-most hand side represents the time integration of a velocity 
and the right-most hand side represents a distance.
We note that $\sqrt{\e^{\Sigma(t)}-1}$ plays the role of velocity.
This is a nonnegative quantity and measures a degree of separation from 
the equilibrium state.
For a given distance,
a large-$t$ is required for a near-equilibrium process, and 
a small-$t$ for a far-from-equilibrium one.
Thus, we have a trade-off relation between time, entropy and state distance.

We note that 
the coefficient $C(t)$ is determined by the ratio 
of second- and third-order fluctuations of $W$.
It highly depends on the system and is a nonuniversal quantity.
On the other hand, 
the average of the work is bounded from below by the second fluctuation
which is related to the universal geometric distance.
We also note that this lower limit is negligible at the thermodynamic limit
since the left-hand side of (\ref{bound}) depends 
linearly on the system size and the right-hand side logarithmically.
The present result is important only for small systems where 
the fluctuation plays a significant role.

We can improve the bound by using the property that the entropy production
is separated into two parts.
It is possible to find a lower bound for each part, 
although the physical meaning is not evident.
$\Sigma_0(t)=D(\hat{\rho}(0\!\to\!t)||\hat{\rho}_{0}(t))$ represents the divergence
for systems without the counterdiabatic term and we obtain 
\be
 \e^{\Sigma_0(t)}-1 \ge \frac{\beta^2}{2}C_0(t)[(W-[W]_t^{(0)})^2]_{t}^{(0)}, 
\ee
where 
\be
 C_0(t)=\exp\left(-\frac{\beta}{3}
 \frac{[(W-[W]_t^{(0)})^3]_{t}^{(0)}}
 {[(W-[W]_t^{(0)})^2]_{t}^{(0)}}\right). 
\ee
The average denoted by $[\ ]_t^{(0)}$ is calculated from the distribution 
\be
 P^{(0)}(W,t)=\sum_n p_n(0)\langle n(t)|
 \delta \left(W-(\hat{H}_0(t)-\epsilon_n(0))\right) |n(t)\rangle,
\ee
This result is derived in the same way as $\Sigma(t)$ 
from the improved Jensen inequality.
We note that the relation $[W]_t=[W]_t^{(0)}$ holds as we see from (\ref{W}).

The bound of $\Sigma_1(t)=D(\hat{\rho}_0(t)||\hat{\rho}(t))$ is 
calculated from the improved Gibbs--Bogoliubov inequality, 
which can be derived from the improved Jensen inequality.
The detail is described in \ref{sec:gb} and we obtain
\be
 \e^{\Sigma_1(t)}-1 \ge 
 \frac{\beta^2}{2}C_1(t)\langle\hat{H}_1^2(t)\rangle_{t}^{(0)}, \label{ineq-s1}
\ee
where 
\be
 C_1(t)=
 \exp\left(-\frac{\beta}{3}\frac{\langle\hat{H}_1^3(t)\rangle_{t}^{(0)}
 -\frac{1}{2}\langle[\hat{H}_1(t),[\hat{H}_1(t),\hat{H}_0(t)]]\rangle_{t}^{(0)}}
 {\langle\hat{H}_1^2(t)\rangle_{t}^{(0)}}\right),
\ee
and the average is with respect to $\hat{H}_0(t)$ as 
\be
 \langle \cdots \rangle_t^{(0)} 
 = \frac{1}{Z_t^{(0)}}\Tr \left(\cdots\right)\e^{-\beta\hat{H}_0(t)}.
\ee
In this case, the second order fluctuation 
in the right-hand side of (\ref{ineq-s1}) is written as 
\be
 \langle\hat{H}_1^2(t)\rangle_{t}^{(0)} 
 =\sum_n p_n(t)\langle \dot{n}(t)|
 \left(1-|n(t)\rangle\langle n(t)|\right)|\dot{n}(t)\rangle,
\ee
where $p_n(t)=\e^{-\beta\epsilon_n(t)}/Z_t^{(0)}$.
which is slightly different from the fluctuation 
in the right-hand side of (\ref{W2c}).
It is not clear whether this quantity is further bounded from below 
by a geometric distance.

From a practical point of view, 
the sum of the lower bounds for $\Sigma_0(t)$ and $\Sigma_1(t)$ 
is expected to be larger than the bound for $\Sigma(t)$ and 
can be a good approximation of the entropy production.
We study simple examples in the next section.

%%%%%%%%%%%%%%%%%%%%%%%%%%%%%%%%%%%%%%%%%%%%%%%%%%%%%%%%%%%%%%%%%%%%%%%%%%%
\subsection{Examples}

First, we consider the two-level system.
As we mentioned in section~\ref{sec:sta}, the Hamiltonian 
$\hat{H}(t)=\hat{H}_0(t)+\hat{H}_1(t)$ is given by 
\be
 \hat{H}(t)=\frac{1}{2}h(t)\bm{n}(t)\cdot\hat{\bm{\sigma}}
 +\frac{1}{2}(\bm{n}(t)\times\dot{\bm{n}}(t))\cdot\hat{\bm{\sigma}},
\ee
where $h(t)$ is positive and $\bm{n}(t)$ is a unit vector.
We note that $h(t)\bm{n}(t)$ corresponds to $\bm{h}_0(t)$ in (\ref{two}).
In the present example, $\Sigma(t)$ is calculated exactly: 
\be
 && \Sigma_0(t)=
  -\frac{\beta(h(t)-h(0))}{2}
 \tanh\left(\frac{\beta h(0)}{2}\right)
 +\ln \left[
 \frac{\cosh\left(\frac{\beta h(t)}{2}\right)}
 {\cosh\left(\frac{\beta h(0)}{2}\right)}
 \right],
 \\
 && \Sigma_1(t) = \ln \left[
 \frac{\cosh\left(\frac{\beta \tilde{h}(t)}{2}\right)}
 {\cosh\left(\frac{\beta h(t)}{2}\right)}
 \right],  
\ee
where $\tilde{h}(t)=\sqrt{h^2(t)+(\bm{n}(t)\times\dot{\bm{n}}(t))^2}$.
$\Sigma_0(t)$ and $\Sigma_1(t)$ are respectively bounded below in the form 
$\Sigma_0(t)\ge \ln (1+\delta_0(t))$ and 
$\Sigma_1(t)\ge \ln (1+\delta_1(t))$.
We note that the calculation of the bound for $\Sigma(t)$ in (\ref{bound})
is cumbersome compared to those of $\Sigma_0(t)$ and $\Sigma_1(t)$.
This is because we need to calculate 
the fluctuations of the total Hamiltonian.
Using the decomposition, we can calculate a bound more easily.

\begin{center}
\begin{figure}[t]
\begin{minipage}[h]{0.45\textwidth}
\begin{center}
\includegraphics[width=1.\columnwidth]{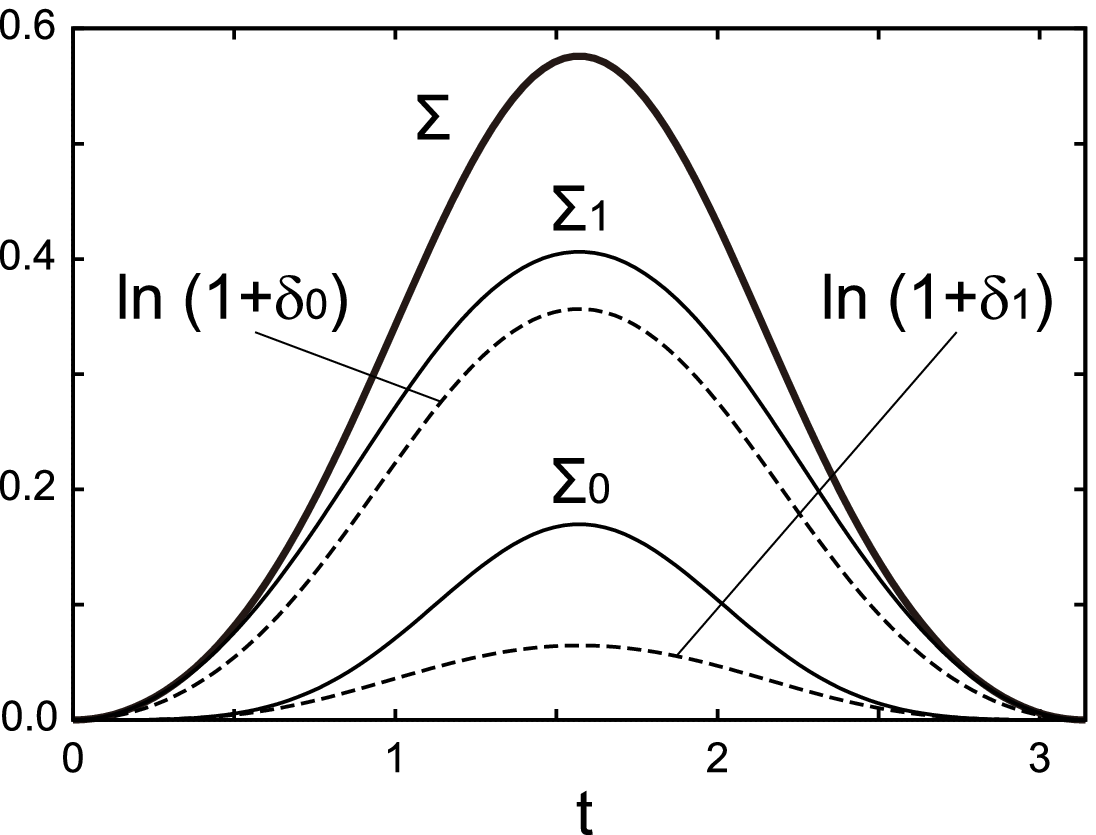}
\caption{$\Sigma(t)=\Sigma_0(t)+\Sigma_1(t)$ for a two-level system.
The result is periodic and is plotted for one period $\pi$.
}
\label{fig:two}
\end{center}
\end{minipage}
\hspace{0.1\textwidth}
\begin{minipage}[h]{0.45\textwidth}
\begin{center}
\includegraphics[width=1.\columnwidth]{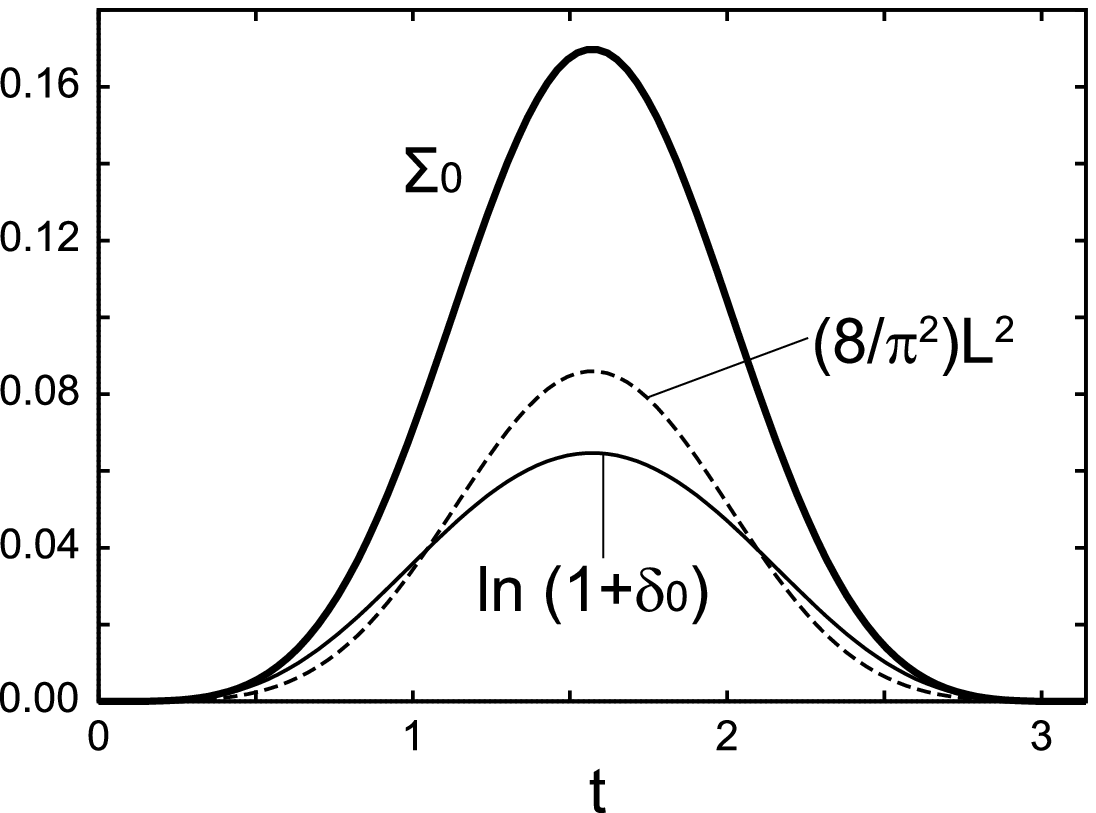}
\caption{$\Sigma_0(t)$ for a two-level system.}
\label{fig:s0-two}
\end{center}
\end{minipage}
\end{figure}
\end{center}

\begin{center}
\begin{figure}[t]
\begin{minipage}[h]{0.45\textwidth}
\begin{center}
\includegraphics[width=1.\columnwidth]{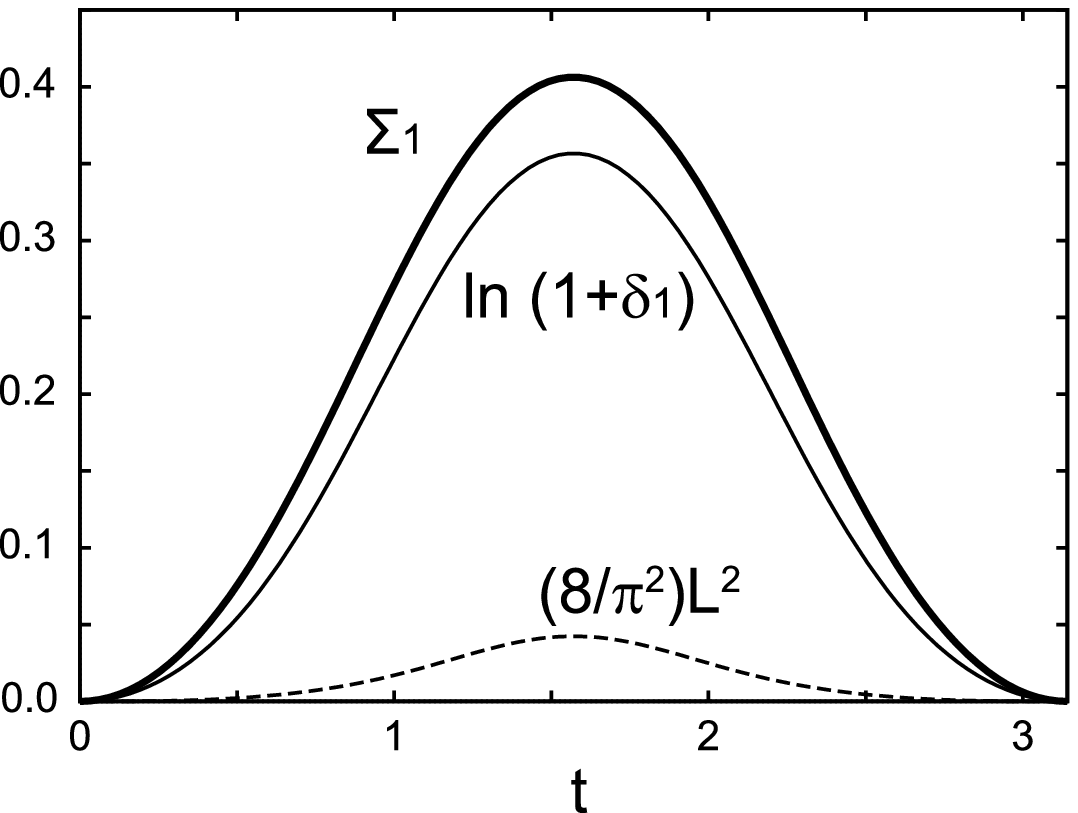}
\caption{$\Sigma_1(t)$ for a two-level system.}
\label{fig:s1-two}
\end{center}
\end{minipage}
\hspace{0.1\textwidth}
\begin{minipage}[h]{0.45\textwidth}
\begin{center}
\includegraphics[width=1.\columnwidth]{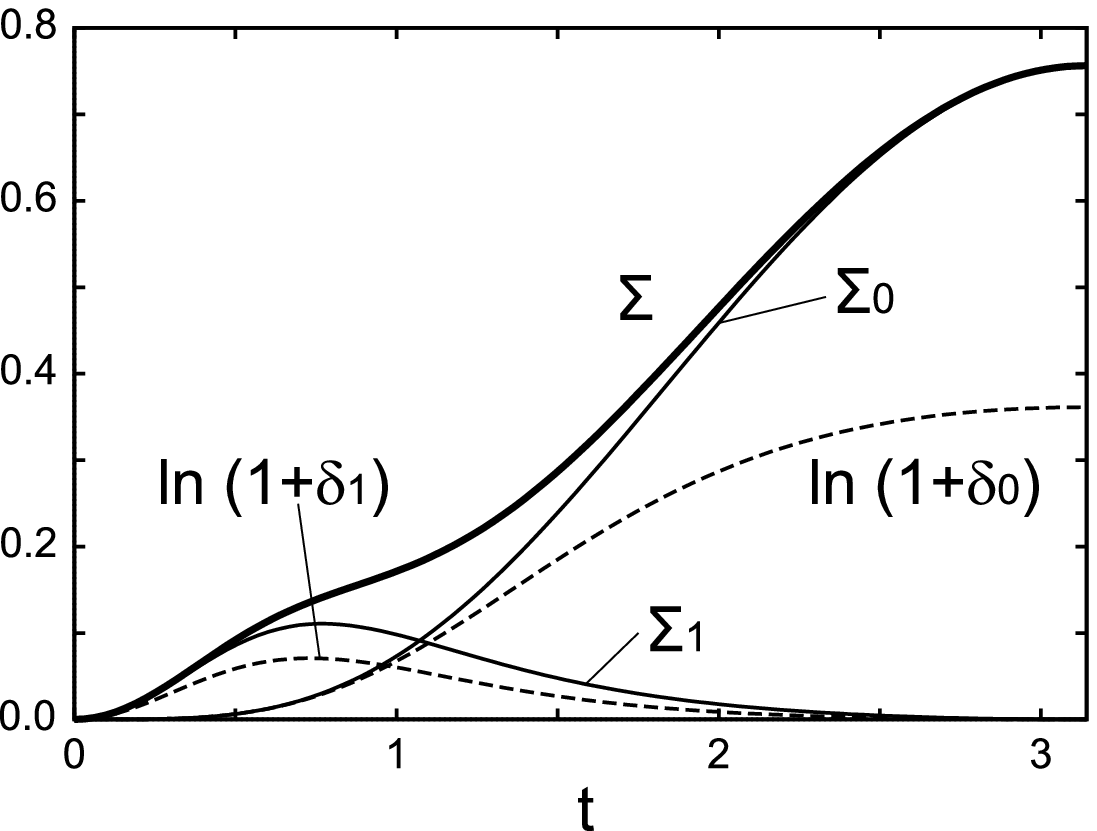}
\caption{$\Sigma(t)=\Sigma_0(t)+\Sigma_1(t)$ for a harmonic oscillator system.
The result is plotted for half period $\pi$.}
\label{fig:ho}
\end{center}
\end{minipage}
\end{figure}
\end{center}

We consider an example 
\be
 && h(t)=1+\frac{1}{2}\cos (2t), \\
 && (\bm{n}(t)\times\dot{\bm{n}}(t))^2=\sin^2 t, 
\ee
with $\beta=2$.
The result is plotted in figure~\ref{fig:two}.
We see that the lower bound gives a good approximation of $\Sigma(t)$
in this example.
We also plot $\Sigma_0(t)$ in figure~\ref{fig:s0-two} 
and $\Sigma_1(t)$ in figure~\ref{fig:s1-two} to compare the result 
with the different bound in (\ref{L2}).
The result shows that our bound becomes better than the bound from (\ref{L2}) 
in some ranges of the parameters, and becomes worse in the other cases.

%%%%%%%%%%%%%%%
%%%%%%%%%%%%
%%%%%%%%%%%%%%%
%%%%%%%%%%%%

The second example is the harmonic oscillator where 
the Hamiltonian is written as 
\be
 \hat{H}(t)=\frac{1}{2m}\hat{p}^2+\frac{m\omega^2(t)}{2}\hat{x}^2
 -\frac{\dot{\omega}(t)}{4\omega(t)}\left(
 \hat{x}\hat{p}+\hat{p}\hat{x}\right),
\ee
where the last term represent the counterdiabatic term~\cite{MCILR}.
$\Sigma_0(t)$ and $\Sigma_1(t)$ are respectively calculated as 
\be
 & &\Sigma_0(t) = 
 \frac{\beta(\omega(t)-\omega(0))}{2\tanh\left(\frac{\beta\omega(0)}{2}\right)} 
 -\ln\left[\frac{\sinh\left(\frac{\beta\omega(t)}{2}\right)}
 {\sinh\left(\frac{\beta\omega(0)}{2}\right)}\right], 
 \\
 & & \Sigma_1(t) = 
 -\ln\left[\frac{\sinh\left(\frac{\beta\Omega(t)}{2}\right)}
 {\sinh\left(\frac{\beta\omega(t)}{2}\right)}\right], 
\ee
where $\Omega(t)=\sqrt{\omega^2(t)-[\dot{\omega}(t)/(2\omega(t))]^2}$.
We note that the condition $\omega^2(t)\ge |\dot{\omega}(t)|/2$ 
is required to make $\Omega(t)$ real.
We parametrize $\omega(t)$ as  
\be
 \omega(t)=1-\frac{1}{2}\cos t, 
\ee
and set $\beta=2$. 
The result is plotted in figure~\ref{fig:ho}.
Again, we can obtain a good approximation of $\Sigma(t)$ using the lower bound.
We see that $\Sigma_1(t)$ is negligible at $t\sim \pi$.
This is because $\dot{\omega}(t)$ takes a small value around the point.

%%%%%%%%%%%%%%%%%%%%%%%%%%%%%%%%%%%%%%%%%%%%%%%%%%%%%%%%%%%%%%%%%%%%%%%%%%%
\section{Conclusions}
\label{sec:conclusion}

In conclusion, we have discussed nonequilibrium properties 
of thermally-isolated systems by using STA.
The main conclusions in this paper are 
(i) STA is applicable to any dynamical systems, (ii)
the entropy production is separated into two parts and 
the information-geometric interpretation is possible
and (iii) the entropy production has a lower limit 
which is used to derive a trade-off relation.

We have stressed that the idea of STA is applicable 
to any nonequilibrium processes.
The property that the Hamiltonian is separated into two parts 
is directly reflected to the nonequilibrium entropy production.
The Pythagorean theorem opens up a novel perspective
in studying the nonequilibrium systems.
Separation of the Hamiltonian can be used not only to solve 
the dynamical problems but also 
to characterize the nonequilibrium properties.
It will also be useful to find an efficient algorithm for a dynamical system.

The lower bound of the entropy production gives 
a new type of trade-off relation between time and 
a notion of distance to equilibrium.
To derive the lower bound, 
we used the improved Jensen inequality.
Although there is no physical meaning of this inequality, 
the lower limit is represented by work fluctuations and is related to 
the geometric distance of two states.
It is interesting to note that the entropy plays a role of velocity.
This can be understood intuitively since the entropy becomes small for 
a quasistatic process where a large time is required to change 
the state to a different one.

Although it is still a difficult problem 
to find the proper separation of a given general Hamiltonian, 
we can invent, for example, a new approximation method 
from an information-geometric point of view.
Actually, the projection theorem is utilized to find an optimized solution
in the problem of information processing.
The present work is only the beginning for 
applications of the concept of the information geometry to 
nonequilibrium dynamics.
We expect that we can find a new efficient algorithm 
based on a picture that we discussed in this paper.

%%%%%%%%%%%%%%%%%%%%%%%%%%%%%%%%%%%%%%%%%%%%%%%%%%%%%%%%%%%%%%%%%%%%%%%%%%%%
\section*{Acknowledgments}
The author is grateful to Ken Funo, Tomoyuki Obuchi 
and Keiji Saito for useful discussions and comments.
This work was supported by JSPS KAKENHI Grant No. 26400385.

%%%%%%%%%%%%%%%%%%%%%%%%%%%%%%%%%%%%%%%%%%%%%%%%%%%%%%%%%%%%%%%%%%%%%%%%%%%%
\appendix
\section{Jensen inequality}
\label{sec:jensen}

For a convex function $f$ of random variables $X$, 
the average satisfies the inequality
\be
 \langle f(X)\rangle \ge f(\langle X\rangle),
\ee
where $\langle\ \rangle$ denotes the average with respect to $X$.
The standard Jensen inequality is obtained by setting $f(X)=\e^X$.
To improve the inequality, Decoster used the convex function~\cite{Decoster} 
\be
 f_N(X)=\e^X-\left(1+X+\frac{X^2}{2!}+\cdots+\frac{X^{2N-1}}{(2N-1)!}\right),
\ee
where $N$ is integer.
The case $N=1$ gives the standard inequality 
$\langle\e^X\rangle\ge \e^{\langle X\rangle}$.
Here we take $N=2$ to improve the inequality.
By using the replacement $X\to X-\langle X\rangle+\alpha$
where $\alpha$ is real, we have  
\be
 \langle\e^{X-\langle X\rangle}\rangle \ge 
 1
 +\e^{-\alpha}\left[\frac{\langle (X-\langle X\rangle)^2\rangle}{2!}
 +\frac{\langle (X-\langle X\rangle)^{3}\rangle
 +3\alpha\langle (X-\langle X\rangle)^2\rangle}{3!}
 \right]. \no\\
\ee
To find the tightest inequality, 
we choose $\alpha$ so that 
the right-hand side of this equation is maximized.
We find $\alpha=-\langle(X-\langle X\rangle)^3\rangle/(3\langle(X-\langle X\rangle)^2\rangle)$ and obtain 
\be
 \langle\e^{X-\langle X\rangle}\rangle \ge 
 1+ \frac{\langle(X-\langle X\rangle)^2\rangle}{2}
 \exp\left(\frac{\langle(X-\langle X\rangle)^3\rangle}
 {3\langle(X-\langle X\rangle)^2\rangle}\right).
 \label{jensen-n3}
\ee

%%%%%%%%%%%%%%%%%%%%%%%%%%%%%%%%%%%%%%%%%%%%%%%%%%%%%%%%%%%%%%%%%%%%%%%%%%%%
\section{Gibbs--Bogoliubov inequality}
\label{sec:gb}

As an application of the Jensen inequality, we consider 
the free energy calculated from the partition function
\be
 Z = \Tr \e^{-\beta H} = \e^{-\beta F}.
\ee
We write the Hamiltonian $H=H_0+H_1$.
Then, using the standard Jensen inequality, we can obtain 
\be
 Z \ge Z_0 \e^{-\beta \langle H_1\rangle_0} 
\ee
where $Z_0 = \Tr \e^{-\beta H_0} = \e^{-\beta F_0}$ and 
$\langle \cdots\rangle = \Tr (\cdots)\e^{-\beta H_0}/Z_0$.
Thus, we have
\be
 F \le F_0+\langle H_1\rangle_0. \label{gb}
\ee
This is the Gibbs--Bogoliubov inequality~\cite{Gibbs, Bogoliubov}.
This result holds for arbitrary separations of $H$.
We also note that the formula holds even when 
$\hat{H}_0$ and  $\hat{H}_1$ do not commute with each other.
When they do not commute, we use the Peierls inequality~\cite{Peierls}   
\be
 \sum_n\langle n|\e^{\hat{X}}|n\rangle
 \ge \sum_n\e^{\langle n|\hat{X}|n\rangle}, 
\ee
where $\hat{X}$ is a hermitian operator and 
$\{|n\rangle\}$ represents a complete basis.

The noncommutativity of the operators becomes important when we consider 
the improved inequality.
The improved Gibbs--Bogoliubov inequality 
corresponding to the improved Jensen inequality in (\ref{jensen-n3}) 
is calculated in a similar way and we obtain 
\be
 F &\le& F_0+\langle H_1\rangle_0 
 -\frac{1}{\beta}
 \ln \left[
 1+\frac{\beta^2}{2}\langle(H_1-\langle H_1\rangle_0)^2\rangle_0
 \right.\no\\
 && \left.
 \times\exp\left(-\frac{\beta}{3}\frac{\langle(H_1-\langle H_1\rangle_0)^3\rangle_0
 -\frac{1}{2}
 \langle[\hat{H}_1,[\hat{H}_1,\hat{H}_0]]\rangle_0}
 {\langle(H_1-\langle H_1\rangle_0)^2\rangle_0}\right)
 \right].
\ee
Since the third term of the right-hand side is negative,
this inequality becomes an improvement of the standard inequality (\ref{gb}).

%%%%%%%%%%%%%%%%%%%%%%%%%%%%%%%%%%%%%%%%%%%%%%%%%%%%%%%%%%%%%%%%%%%%%%%%%%%%
%%%%%%%%%%%%%%%%%%%%%%%%%%%%%%%%%%%%%%%%%%%%%%%%%%%%%%%%%%%%%%%%%%%%%%%%%%%%

\end{document}